\documentclass[twocolumn,twoside,letterpaper,11pt]{article}
\usepackage{graphicx,xrrc,natbib}
\usepackage{times}

\begin{document}

\title{XMM-NEWTON OBSERVATIONS OF M31: X-RAY PROPERTIES OF RADIO SOURCES AND SNR CANDIDATES}



\author{    S. Trudolyubov} 
\institute{ Institute of Geophysics and Planetary Physics, University of California, Riverside} 
\address{   1432 Geology Building, Riverside, CA 92521, U.S.A.} 
\email{     sergeyt@citrus.ucr.edu                         } 

\author{    W. Priedhorsky}
\institute{ Los Alamos National Laboratory}
\address{   LANL, Los Alamos, NM 87545, U.S.A.} 
\email{     wpriedhorsky@lanl.gov}


\maketitle

\abstract{We present the results of the ongoing {\em XMM-Newton} Survey of nearby spiral galaxy M31. 
17 X-ray sources detected in the survey have bright radio counterparts, and 15 X-ray sources coincide 
with supernova remnant (SNR) candidates from optical and radio surveys.

15 out of 17 sources with radio counterparts, not SNR candidates, have spectral properties similar to 
that observed for background radio galaxies/quasars or Crab-like supernova remnants located in M31. The 
remaining two sources, XMMU J004046.8+405525 and XMMU J004249.1+412407, have soft X-ray spectra, and 
are associated with spatially resolved H-alpha emission regions, which makes them two new SNR candidates 
in M31.

The observed absorbed X-ray luminosities of SNR candidates in our sample range from $\sim 10^{35}$ to 
$\sim 5 \times 10^{36}$ ergs s$^{-1}$, assuming the distance of 760 kpc. Most of the SNR candidates 
detected in our survey have soft X-ray spectra. The spectra of the brightest sources show presence of 
emission lines and can be fit by thermal plasma models with $kT \sim 0.1 - 0.4$ keV. The results of 
spectral fitting of SNR candidates suggest that most of them should be located in a relatively low 
density regions.

We show that X-ray color-color diagrams can be useful tool for distinguishing between intrinsically 
hard background radio sources and Crab-like SNR and thermal SNR in M31 with soft spectra.
}

\section{Introduction}
The Andromeda Galaxy (M31), the closest giant spiral galaxy to our own, is a unique object for 
the study of optical and X-ray astronomy. Its proximity and favorable orientation allow to observe 
stellar populations over the full extent of the galaxy at a nearly uniform distance, and with less 
severe effects of line-of-sight contamination from interstellar gas and dust. Due to similarities 
between the two galaxies, the results from the study of M31 provide an important benchmark for 
comparison with the results from the study of our own Milky Way Galaxy. M31 was observed extensively 
in X-rays with {\em Einstein}, {\em ROSAT}, {\em Chandra} and {\em XMM} missions, detected hundreds 
of sources, with some bright X-ray sources coincident with radio-emitting sources and SNR candidates 
from optical surveys. In the extensive {\em ROSAT}/PSPC survey of M31 (Supper et al. 2001), 16 X-ray 
emitting SNR were identified. Recently, Kong et al. (2003) identified two new SNRs using the data of 
{\em Chandra} and {\em VLA} observations. Trudolyubov et al. (2004) used {\em XMM-Newton} observations 
to study X-ray spectral properties of the 3 bright thermal SNR in the northern disk of M31.   

\section{Source identification and classification}
Using the data of {\em XMM-Newton} survey of M31, we detected about 600 X-ray point sources in the six 
{\em XMM} fields covering major axis of M31. We searched for radio, optical and X-ray counterparts to 
the XMM sources using the existing catalogs and images from the CTIO/KPNO Local Group Survey (LGS) 
(Massey et al. 2001). We varied the search radius based on both the accuracy of the catalogs and 
localization errors of {\em XMM} sources (upper limit of 5 arcseconds). We used the following catalogs 
and corresponding search radii:

\noindent {\em Radio sources:} VLA All-sky Survey Catalog (Condon et al. 1998) and the lists of Walterbos, 
Brinks $\&$ Shane (1985) and Braun (1990) -- 5 arcsecond search radius. 17 X-ray sources were found to 
coincide with radio sources.

\noindent {\em Supernova remnant candidates:} the lists by Braun \& Walterbos (1993) and Magnier et al. 
(1995) -- 10 arcsecond search radius. 15 X-ray sources have SNR counterparts in M31.

\noindent {\em X-ray sources:} the {\em ROSAT}/PSPC catalog of sources in the field of M31 (Supper et al.
2001) -- search radius specified by position accuracy for each individual source.

\begin{figure}
\begin{center}
\includegraphics[width=7.9cm]{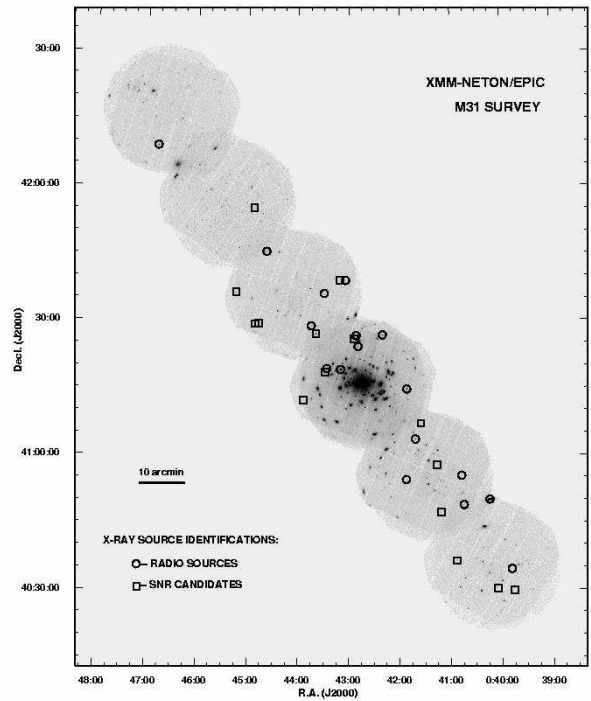}
\caption{\small {\em XMM}/EPIC X-ray image of M31 in the $0.3 - 10$ keV energy range. X-ray sources 
identified with radio sources and SNR candidates are marked by {\em circles} and {\em squares}.}
\label{image_epic_all}
\end{center}
\end{figure}

The information on the positions and identifications of the X-ray sources coincident with radio sources 
and SNR candidates is shown in Tables \ref{radio_ID} and \ref{SNR_ID}. The {\em XMM}/EPIC X-ray images of 
M31 with source positions marked are shown in Fig. \ref{image_epic_all}.

\begin{table}[h]
\begin{center}
\caption{X-ray sources coincident with radio sources not identified as SNR}\medskip
\label{radio_ID}
\small
\begin{tabular}{lll}
Source Name   & Flux$^{a}$    & Radio/X-ray ID$^{b}$\\
XMMU J00...   &               &                     \\
\hline\hline
3948.2+403436 & $0.19\pm0.04$ & 37W42               \\
4013.9+405004 & $856.6\pm7.5$ & 37W51               \\
              &               & SHP67               \\
              &               & BL Lac              \\
4044.0+404853 & $0.17\pm0.03$ & 37W63               \\
              &               & SHP90               \\
4046.8+405525 & $0.72\pm0.11$ & 37W64               \\
              &               & SHP94               \\
              &               & SNR?                \\
4141.2+410331 & $7.53\pm0.52$ & 37W91               \\
              &               & SHP119              \\
4151.4+411439 & $9.31\pm0.51$ & 37W95               \\
              &               & SHP132              \\
4152.0+405429 & $0.12\pm0.03$ & 37W97               \\
4220.2+412641 & $1.02\pm0.05$ & 37W116              \\
              &               & RX J0042.3+4126     \\
4249.1+412407 & $1.24\pm0.09$ & 37W140              \\
              &               & SNR?                \\
4251.4+412633 & $0.22\pm0.04$ & 37W144              \\
4304.1+413848 &               & 37W150              \\
4309.8+411900 & $0.11 - 52.91$& 37W153              \\
              &               & SHP226              \\
              &               & Highly variable     \\
4326.2+411912 & $0.23\pm0.04$ & 37W158B             \\
4329.3+413554 & $0.17\pm0.03$ & 37W159              \\
4344.6+412843 & $1.80\pm0.20$ & 37W169              \\
4437.8+414513 & $7.32\pm0.61$ & 37W194              \\
4648.0+420852 & $10.04\pm0.44$& 37W235              \\
              &               & SHP353              \\
\hline
\end{tabular}

\begin{list}{}
\item $^{a}$ -- Source X-ray flux in the $0.3 -10$ keV energy band in units of 
$10^{-14}$ ergs s$^{-1}$ cm$^{-2}$
\item $^{b}$ -- Source identifications beginning with 37W refer to the radio sources listed 
in Walterbos, Brinks $\&$ Shane (1985). Identifications beginning with SHP refer to entries 
from {\em ROSAT}/PSPC catalog of X-ray sources in M31 by Supper et al. (2001).
\end{list}

\end{center}
\end{table}

\begin{table}[h]
\begin{center}
\caption{X-ray sources identified with SNR candidates}\medskip
\label{SNR_ID}
\small
\begin{tabular}{lcl}
Source Name    & $L_{\rm X}^{a}$(0.3-10 keV)     & Identification$^{b}$\\
XMMU J00...      & $\times 10^{35}$ ergs s$^{-1}$&                     \\
\hline\hline
3945.1+402950 & $1.74\pm0.28$&MA 1-001\\
4005.1+403017 & $3.31\pm0.90$&MA 2-002\\
4052.5+403626 & $7.22\pm1.36$&MA 3-027\\
4110.6+404716 & $0.82\pm0.20$&MA 2-013\\
4115.6+405741 & $1.45\pm0.36$&MA 2-016\\
4134.8+410659 & $6.43\pm0.83$&BW53\\
4253.6+412551 &$34.47\pm1.37$&BW55\\
4310.6+413853 & $0.96\pm0.21$&BW6\\
4327.9+411828 &$48.26\pm1.14$&BW56,MA 2-043\\
4339.1+412655 &$18.98\pm0.74$&BW57,MA 3-069\\
4353.7+411206 & $6.22\pm1.13$&BW9\\
4447.1+412917 & $1.07\pm0.25$&BW29\\
4451.1+412907 & $9.30\pm0.80$&BW33\\
4452.7+415457 & $2.02\pm0.51$&BW34\\
4514.0+413614 & $9.13\pm0.70$&BW39,MA 2-048\\
\hline
\end{tabular}

\begin{list}{}
\item $^{a}$ -- X-ray luminosity of the source assuming a distance of 760 kpc.
\item $^{b}$ -- Source identifications beginning with BW refer to 
the SNR candidates listed in  Braun $\&$ Walterbos (1993). Identifications 
beginning with MA refer to the SNR candidates from Magnier et al. (1995). 
\end{list}

\end{center}
\end{table}

\section{Radio Sources}

\begin{figure}
\begin{center}
\includegraphics[width=\columnwidth]{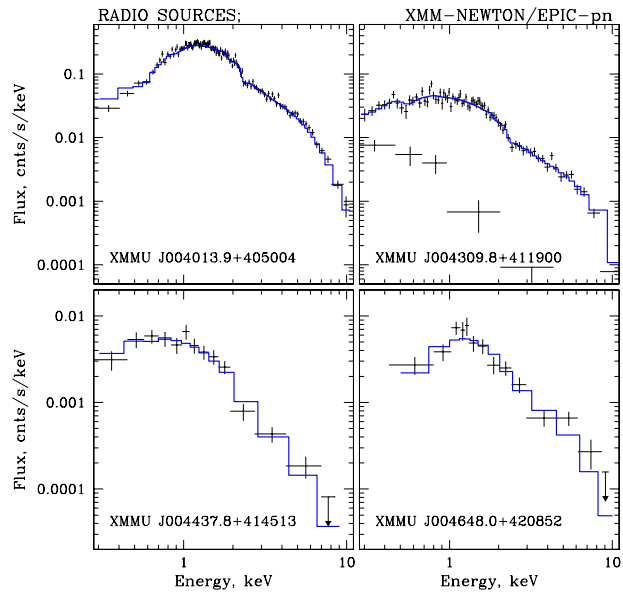}
\caption{\small EPIC-pn count spectra of the four brightest X-ray sources coincident with radio sources. 
For each spectrum the absorbed power law model fits are shown with {\em blue solid} histograms. The spectra 
of variable source XMMU J004309.8+411900, corresponding to the high and low levels of X-ray flux are shown 
in the {\em upper right} panel.}
\label{RS_spec_image}
\end{center}
\end{figure}

17 bright {\em XMM} sources not identified as SNR candidates coincide with bright 
radio sources detected with VLA all-sky (Condon et al. 1998) and earlier surveys (Walterbos, Brinks $\&$ 
Shane 1985; Braun 1990). 15 out of 17 sources with radio counterparts have spectral properties similar 
to that observed for background radio galaxies/quasars and Crab-like SNR in M31. The brightest objects have 
X-ray spectra well presented by absorbed power laws with indices between $\sim 1.6$ and $\sim 2.2$, typical 
for this source class (Table \ref{rs_spec_fit}; Fig. \ref{RS_spec_image}). For majority of bright radio 
sources the corresponding values of the interstellar absorption inferred from their X-ray spectra are well 
above expected Galactic value in the direction of M31 (Dickey $\&$ Lockman 1990), supporting their 
identification with objects in the background of M31.

\begin{figure}
\begin{center}
\includegraphics[width=\columnwidth]{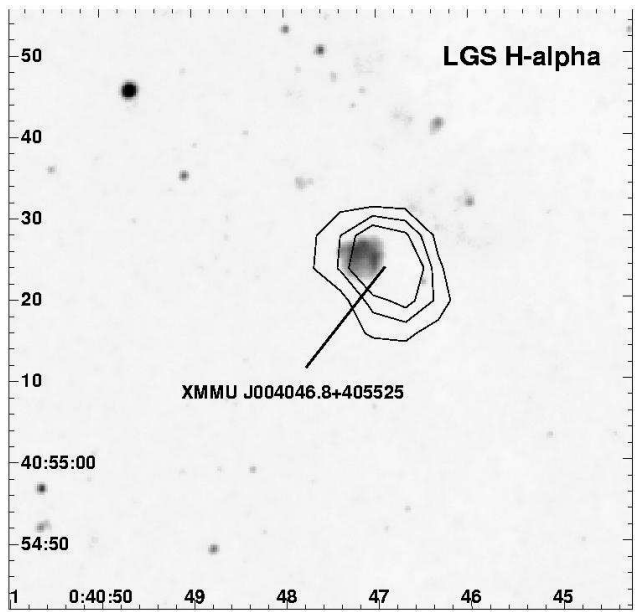}
\caption{\small Optical H$_{\alpha}$ image of the southern disk of M31 from the Local Group Survey 
(Massey et al. 2001) with {\em XMM}/EPIC X-ray contours overlaid. The image shows spatially resolved 
H$_{\alpha}$ region, associated with soft X-ray source XMMU J004046.8+405525 and a radio sources 37W64 
from Walterbos, Brinks and Shane (1985) and GLG068 from Gelfand et al. (2004).}
\label{SNR_candidate_1_image}
\end{center}
\end{figure}

\begin{figure}
\begin{center}
\includegraphics[width=\columnwidth]{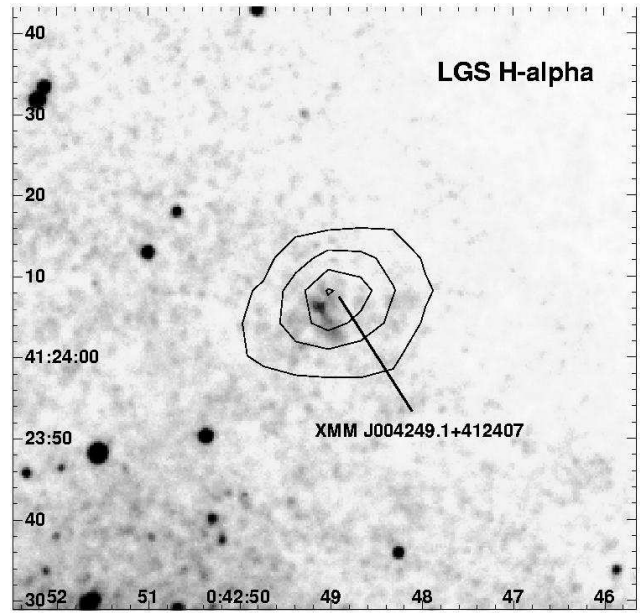}
\caption{\small Optical H$_{\alpha}$ image of the southern disk of M31 from the Local Group Survey 
(Massey et al. 2001) with {\em XMM}/EPIC X-ray contours overlaid. The image shows spatially resolved 
H$_{\alpha}$ region, associated with soft X-ray source XMMU J004249.1+412407 and a radio source 37W140 
from Walterbos, Brinks and Shane (1985).}
\label{SNR_candidate_2_image}
\end{center}
\end{figure}

Two {\em XMM} sources, identified with radio-emitting objects, have soft X-ray spectra. The X-ray spectrum 
of the first source, XMMU J004309.8+411900, can be well fit to the absorbed Raymond-Smith (RS) thermal plasma 
emission model with $kT \sim 0.2$ keV. The corresponding absorbed source luminosity in the $0.3 - 3$ keV 
energy band is $\sim 5 \times 10^{35}$ ergs s$^{-1}$ assuming the distance of 760 kpc. The other source, 
XMMU J004249.1+412407, has a spectrum well approximated by the absorbed RS model with $kT \sim 0.4$ keV and 
luminosity of $\sim 9 \times 10^{35}$ ergs s$^{-1}$ in the $0.3 - 3$ keV energy range. The inspection of 
the optical LGS images of M31 revealed two spatially resolved H$_{\alpha}$ emission regions, coincident 
with these two X-ray sources (Fig. \ref{SNR_candidate_1_image}, \ref{SNR_candidate_2_image}). The combination 
of the soft thermal X-ray spectra, extended H$_{\alpha}$ and bright radio counterparts allows to propose 
XMMU J004309.8+411900 and XMMU J004249.1+412407 as a new SNR candidates in M31. 

\begin{table}[h]
\begin{center}
\caption{\small Spectral fit results for six brightest X-ray sources identified with radio-emitting 
objects ({\em XMM}/EPIC data, absorbed simple power law model, $0.3 - 10.0$ keV energy 
range}\medskip
\label{rs_spec_fit}
\small
\begin{tabular}{lccc}
Source Name    & Photon & N$_{\rm H}^{a}$ & Flux$^{b}$\\
XMMU J00...    & Index  &                 &           \\
\hline\hline
4013.9+405004 &$2.05\pm0.03$         &$0.47\pm0.01$         &$85.66\pm0.75$\\
4141.2+410331 &$1.60\pm0.13$         &$0.15^{+0.05}_{-0.04}$&$0.75\pm0.05$ \\
4151.4+411439 &$1.89^{+0.15}_{-0.13}$&$0.17\pm0.04$         &$0.93\pm0.05$ \\
4309.8+411900 &$1.80\pm0.04$         &$0.15\pm0.01$         &$5.29\pm0.09$\\
4437.8+414513 &$2.15^{+0.11}_{-0.13}$&$0.17^{+0.03}_{-0.04}$&$0.73\pm0.06$ \\
4648.0+420852 &$1.86^{+0.10}_{-0.12}$&$0.40^{+0.10}_{-0.05}$&$1.00\pm0.04$ \\
\hline
\end{tabular}

\begin{list}{}
\item $^{a}$ -- Equivalent hydrogen absorbing column in units of $10^{22}$ cm$^{-2}$.
\item $^{b}$ -- Source X-ray flux in the $0.3 -10$ keV energy band in units of 
$10^{-13}$ ergs s$^{-1}$ cm$^{-2}$
\end{list}

\end{center}
\end{table}

\section{Supernova Remnant Candidates}
15 {\em XMM} sources detected in M31 fields were identified with SNR candidates from optical surveys 
(Braun $\&$ Walterbos 1993; Magnier et al. 1995). Most of these sources were previously detected with 
{\em ROSAT} and identified as SNR (Supper et al. 2001; Magnier et al. 1997).

\begin{figure}
\begin{center}
\includegraphics[width=\columnwidth]{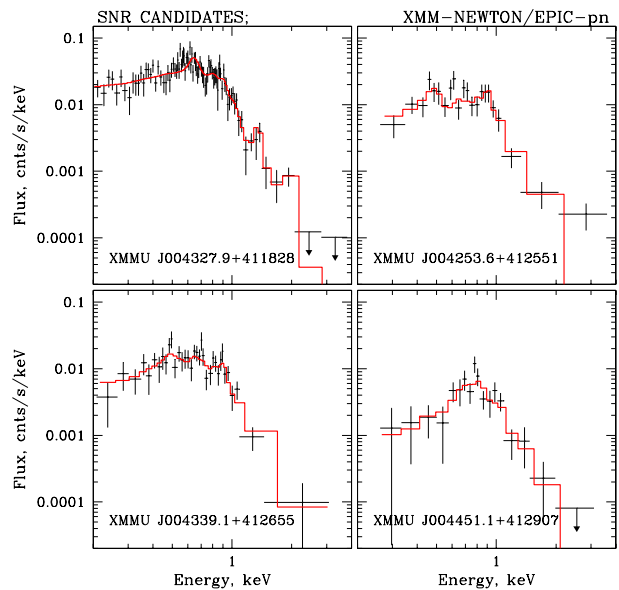}
\caption{\small EPIC-pn count spectra of four brightest SNR candidates in the $0.3 - 3.0$ keV 
energy band. For each spectrum the absorbed Raymond-Smith thermal plasma model fits are shown 
with {\em red solid} histograms.}
\label{SNR_spec_image}
\end{center}
\end{figure}

All relatively bright SNR candidates have relatively soft X-ray spectra (Fig. \ref{SNR_spec_image}). 
We fitted the spectra 
of SNR candidates with various single component spectral models including a simple power law, thermal 
bremsstrahlung, black body, RS thermal plasma, and non-equilibrium ionization collisional plasma (NEI) 
models with interstellar absorption. RS models with characteristic temperatures of $0.1 - 0.4$ keV and 
NEI models with temperatures of $0.5 - 1.5$ keV give the best approximation to the data for 6 bright SNR 
candidate sources (Fig. \ref{SNR_spec_image}). The absorbed X-ray luminosities of SNR candidates range 
from $\sim 10^{35}$ to $\sim 5 \times 10^{36}$ ergs s$^{-1}$ assuming the distance of 760 kpc. The 
estimated ages of SNR range from $\sim 2000$ to $20000$ years and the number density of ambient gas is 
$\sim 0.005-0.4$ cm$^{-3}$, suggesting that these SNR may be located in a relatively low density regions. 
The X-ray measured chemical abundances were found to be similar to their optically determined values 
(Blair, Kirshner \& Chevalier 1982).  

\section{X-ray Color-color Diagram}

In order to facilitate comparison between spectral properties of the radio sources and SNR candidates 
detected in our survey , we constructed their X-ray color-color diagram. We calculated the total number 
of counts for each source using its corrected EPIC-pn spectra in three energy bands: the soft band 
($0.3 - 1.0$ keV), medium band ($1.0 - 2.0$ keV) and hard band ($2.0 - 7.0$ keV). Two X-ray colors were 
defined for each source as: $HR1 = (S - M)/T$ (soft color) and $HR2 = (H - M)/T$ (hard color), where 
$S, M,$ and $H$ are the counts in soft, medium and hard bands respectively, and $T$ is the total number 
of source counts in the $0.3 - 7.0$ keV energy range. We used the data for 16 brightest (10 radio sources 
and 6 SNR candidates) sources, each of which provided more than 50 counts in EPIC-pn.

\begin{figure}
\begin{center}
\includegraphics[width=\columnwidth]{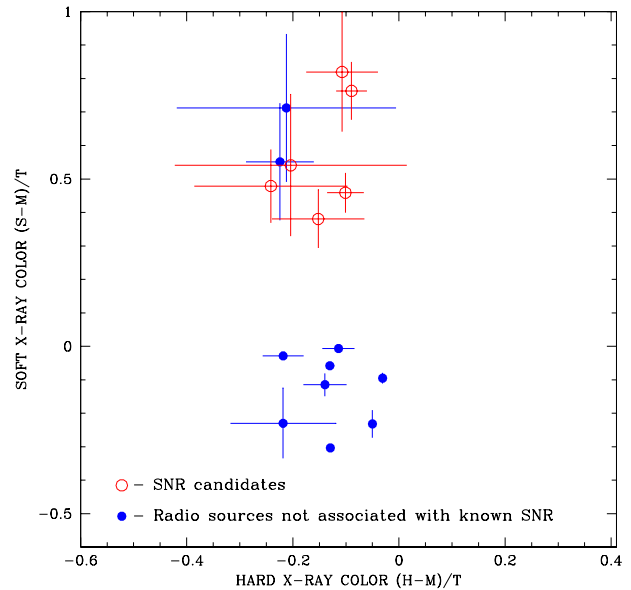}
\caption{\small X-ray color-color diagram of the brightest radio sources ({\em blue filled circles}) and 
SNR candidates ({\em red open circles}).}
\label{colors}
\end{center}
\end{figure}

Fig. \ref{colors} shows the X-ray color-color diagram for the bright X-ray sources identified with radio 
sources and SNR candidates. There are two distinct concentrations of sources in this diagram. The first 
group with $HR1 > 0.3$ includes intrinsically soft sources: bright thermal SNR and two radio sources, 
XMMU J004046.8+405525 and XMMU J004249.1+412407. The second group includes harder radio sources. The 

X-ray color-color diagrams show that they can be useful tool for distinguishing between sources with 
intrinsically soft and hard spectra like thermal SNR and background radio sources. On the other hand, this 
method shows no difference between Crab-like SNR and background radio sources. The combination of 
low-energy absorption and limited instrument bandpass also have significant effect on the source position 
on the color-color diagram (Di Stefano $\&$ Kong 2003). For example, the source with intrinsically soft 
spectrum, if highly absorbed, can easily ``migrate'' to the region on the color diagram normally occupied 
by sources with much harder spectra. Therefore, additional information, such as X-ray variability, 
luminosity and source counterparts at other wavelengths, is needed to classify the X-ray source.

\section{Summary}
Nearly 600 X-ray sources were detected in the ongoing {\em XMM-Newton} survey of M31. 17 of them have 
bright radio counterparts not identified with SNR candidates, and 15 sources coincide with SNR candidates 
from radio and optical surveys.

15 out of 17 sources with radio counterparts, not known to be SNR candidates from optical surveys, have 
spectral properties similar to that observed for background radio galaxies/quasars or Crab-like supernova 
remnants located in M31. The energy spectra of the brightest objects are well represented by absorbed 
power law with photon indices between $\sim 1.6$ and $\sim 2.2$, and low-energy absorption in excess 
of the Galactic foreground value in the direction of M31. The remaining two sources, XMMU J004046.8+405525 
and XMMU J004249.1+412407, have soft X-ray spectra, and are associated with spatially resolved H-alpha 
emission regions, which makes them two new SNR candidates in M31.

Most of the SNR candidates detected in our survey have soft X-ray spectra. The spectra of the brightest 
sources show presence of emission lines and can be fit by thermal plasma models with $kT \sim 0.1 - 0.4$ 
keV. The observed absorbed X-ray luminosities of our SNR candidates range from $\sim 10^{35}$ to 
$\sim 5 \times 10^{36}$ ergs s$^{-1}$, assuming the distance of 760 kpc. The results of spectral fitting 
suggest that most of the bright SNR candidates in our sample may be located in the low density regions. 

We show that X-ray color-color diagrams can be useful tool for distinguishing between intrinsically hard 
background radio sources and Crab-like SNR and thermal SNR in M31.  

\section*{Acknowledgments}
Support for this work was provided through NASA {\em XMM} Grant NAG5-12390. {\em XMM-Newton} is an ESA 
Science Mission with instruments and contributions directly funded by ESA Member states and the USA 
(NASA). This research has made use of data obtained through the High Energy Astrophysics Science Archive 
Research Center Online Service, provided by the NASA/Goddard Space Flight Center.

\end{document}